\documentclass[preprint,final,12pt]{elsarticle}

\usepackage{latexsym,nomencl,amsmath,amssymb}
\usepackage{graphicx}
\journal{Wave Motion}

\begin{document}

\begin{frontmatter}

\title{Wave dispersion under finite deformation}

\author[abdn]{Mohammad H. Abedinnasab}

\author[mih]{Mahmoud I. Hussein\corref{cor1}}
 \ead{mih@colorado.edu}

\address[abdn]{Department of Mechanical Engineering, Sharif University of Technology, Tehran, 11365-9567, Iran}
\cortext[cor1]{Corresponding author. Tel.: +1 303-492-3177}
\address[mih]{Department of Aerospace Engineering Sciences, University of Colorado Boulder, Boulder, CO 80309-0429, USA}

\begin{abstract}
We derive exact dispersion relations for axial and flexural elastic
wave motion in a rod and a beam under finite deformation. For axial
motion we consider a simple rod model, and for flexural motion we
employ the Euler-Bernoulli kinematic hypothesis and consider both a
conventional transverse motion model and an inextensional planar
motion model. The underlying formulation uses the Cauchy stress and the Green-Lagrange strain without omission of higher order terms. For all models, we consider linear constitutive relations in order to isolate the effect of finite motion. The proposed methodology, however, is applicable to problems that also exhibit material nonlinearity. For the rod model, we obtain the exact analytical explicit solution of the derived finite-deformation dispersion relation, and compare it with data obtained via numerical simulation of nonlinear wave propagation in a finite rod. For the beam model, we obtain an approximate solution by standard root finding. The results allow us to quantify the deviation in the dispersion curves when exact large deformation is considered compared to following the assumption of infinitesimal
deformation. We show that incorporation of finite deformation following the chosen definitions of stress and strain raises
the frequency branches for both axial and flexural waves and
consequently also raises the phase and group velocities above the
nominal values associated with linear motion. For the beam problem,
only the inextensional planar motion model provides an accurate
description of the finite-deformation response for both static
deflection and wave dispersion; the conventional transverse motion
model fails to do so. Our findings, which represent the first derivation of finite-strain, amplitude-dependent dispersion relations for any type of elastic media, draw attention to (1) the tangible effect of finite deformation on wave dispersion and consequently on the speed of sound in an elastic medium and (2) the importance of incorporating longitudinal-transverse motion coupling in both
static and dynamic analysis of thin structures subjected to large
nonlinear deformation.
\end{abstract}

\end{frontmatter}

\section{Introduction} \label{Int}

The dispersion relation provides a fundamental characterization of
the nature of wave propagation in an elastic solid. Its derivation
for various solid configurations, such as beams, plates, surfaces
etc., has been key to the development of the field of elastic wave
propagation, a field that traces its beginnings to the seminal
memoir of Poisson \cite{Poisson_1829}. In this memoir, and in the
work of Cauchy \cite{Cauchy_1830}, it was revealed that two types of
waves exist in solids: longitudinal and transverse. Analysis of the
dispersive nature of these waves in different types of unbounded
structures has been a focus of many classical studies since then.
Brief surveys on the historical development of theoretical elastic wave propagation research are provided by Graff \cite{Graff} and Ben-Menahem
\cite{Ben-Menahem_1995}, and a thorough discussion on the reconciliation of theory with experiments on mechanical waves is given by Thurston \cite{Thurston84}. 

These early investigations as well as the vast majority of contemporary studies of wave propagation in elastic solids are based primarily
on linear analysis, that is, linear constitutive laws and linear
strain-displacement relationships are assumed (see \cite{Graff},
\cite{Achenbach}, and references therein). The incorporation of nonlinear effects has nevertheless been actively pursued and continues to attract attention as it allows for a more accurate description
of the underlying motion and facilitates the study of amplitude-dependent wave interaction phenomena that do not appear in linear systems (see the monograph on nonlinear oscillations by Nayfeh and Mook  \cite{Nayfeh_Mook} and a recent special journal issue edited by Destrade and Saccomandi \cite{Destrade_Saccomandi_2009} for broad listings of references on the subject). From an engineering perspective, the effects of nonlinearity on the dispersion of waves in waveguides could be utilized to enrich the design of materials and structural components. In general, the study of nonlinear elastic wave propagation is relevant to nonlinear vibration analysis \cite{Shaw_1993,Kerschen_2006}, dislocation and crack dynamics analysis
\cite{Gumbsch_1999,RosakisAJ_1999}, geophysical and seismic motion
analysis \cite{Sen_1991,Ostrovsky_2001}, material characterization
and nondestructive evaluation \cite{Zheng_1999,VanDenAbeele_2000}, biomedical imaging \cite{Ward_1997}, and others.

Finite amplitude wave propagation in elastic solids is a subset
among the broader class of nonlinear wave propagation problems. From a mathematical perspective, a formal treatment of finite deformation requires the incorporation of an exact nonlinear strain tensor in setting up the governing equations of motion \cite{Ogden}. As a result, the emerging analysis should permit large and finite strain fields, as opposed to small and infinitesimal strain fields. A large portion of research on finite amplitude waves considers initially strained materials (see, for example, the early studies by Truesdell \cite{Truesdell1961} and Green \cite{Green1963} for extensive discussions on the topic). Among the relatively recent works that focused on finite-amplitude plane waves in materials subjected to a large static finite deformation include those of Boulanger and Hayes
\cite{Boulanger_Hayes_1992}, Boulanger et al.
\cite{Boulanger_1994} and Destrade and Saccomandi
\cite{Destrade_Saccomandi_2008}. Furthermore, analysis of finite-amplitude waves in solids often involve small parameters or asymptotic expansions (see Norris \cite{Norris_1999} for a review). Auld \cite{Auld}, for example, and subsequent studies by de Lima and Hamilton \cite{deLimaWJN_2003}, Deng \cite{Deng_2003} and Srivastava et al. \cite{Srivastava_2010}, treated the problem of finite-strain waves using normal mode expansion and forced response calculations. This approach is based on perturbation theory and is therefore limited to small amplitudes. Focusing on rods, modeled in one dimension or higher, many studies considered finite amplitude waves for both incompressible and compressible materials (e.g.,
\citep{Bhatnagar,Thurston84,Wright_1985,Coleman_Newman_1990,Samsonov_1994,DaiHH_1998,Dai_Hu_2000,Porubov_2003}). Of particular relevance is a recent investigation by Zhang and Liu
\cite{Zhang_Liu} in which an exact equation of motion for a rod and an approximate equation of motion for an Euler-Bernoulli beam were
derived under the condition of finite deformation. It is rather common in the study of structural waveguides to omit (for simplicity) some terms in the strain tensor at the outset, and among the strain tensor terms that are retained some emerging high order terms associated with the deformation of the cross-sectional frame are often neglected.

A broad overview of the listed references and other finite-deformation-based studies in the literature reveals that generally the interest has been in obtaining spatial/temporal solutions, or solutions at certain physical limits, rather than complete dispersion relations that directly relate frequency to wavevectors (or wavenumbers). In this paper, we provide an exact analysis of static deflection and elastic wave dispersion in a rod and a beam under finite
deformation with no modeling limitation on the amplitude of the
deflection or the travelling wave. Starting with Hamilton's
principle we consider both axial deformation (to represent
longitudinal motion in a rod) and flexural deformation (to represent
transverse or longitudinal-transverse motion in a beam). Focusing on homogeneous
waveguides with constant cross-section and non-dissipative, isotropic material properties, we derive the exact equation of motion and dispersion relation.\footnote[1]{A derivation of finite-strain dispersion for a beam model based on an approximate equation of motion was published earlier by the authors \cite{Abedinnasab_IMECE_2011}} For the rod problem, we subsequently obtain the explicit frequency versus wavenumber solution. For the flexural beam problem, we obtain a solution by numerically finding the roots of the derived dispersion relation. In all our derivations, we use the Cauchy and Green-Lagrange definitions of stress and strain, respectively. Furthermore, all terms in the nonlinear strain tensor are retained and no high order terms emerging from the differentiation are subsequently neglected. In order to isolate the effect of finite motion, we consider linear  constitutive relations. The proposed methodolgy, however, is applicable to problems that also exhibit material nonliearity. To validate our theoretical approach, we first consider static deflection, in both a
rod and a beam, and make comparisons with finite element (FE) solutions
based on exact elasticity theory. We then consider the dynamics
regime, where for the rod problem we examine wave propagation in a
finite rod by means of standard finite-strain numerical simulations
(also using FE analysis) and compare the full-spectrum response with
our derived dispersion relation. 

The dispersion formulations we develop in this
work, which represent the first derivation of
finite-strain, amplitude-dependent dispersion relations for any type of elastic media, allow us to examine (1) the effect of finite deformation on the
frequency and phase/group velocity dispersion curves in rods and beams and
(2) the role that longitudinal-transverse motion coupling plays in the
underlying nonlinear mechanics of flexural beams.


\section{Rod and beam kinematics based on finite deformation} \label{Kin}

\subsection{Classification of rod and beam kinematics}

In this paper we study geometrically nonlinear wave propagation in a
rod and in a flexural beam, assuming slender structure for each. The rod case is also equivalent to the problem of propagation of plane longitudinal waves along a single direction in a medium in which no lateral boundaries are imposed [i.e., in the linear regime, the model represents a nondispersive one-dimensional (1D) medium]. To put the kinematical description of the problems we consider in context, we show in Table \ref{Delta} the categories of a class of theories for the treatment of a rod/beam, ranging from $1$ to $4$ degree-of-freedom (DOF) systems. The table lists longitudinal motion for a rod (one model), transverse motion for a beam (one model), planar motion for a beam (two models) and spatial motion for a beam (two
models). Each model is described fundamentally by the number and type of
variables, the form of the exact displacement field, $\Delta$, and
the constraint (if any) that applies to the displacement field
variables.

\begin{figure*}[!ht]
\centerline{\includegraphics{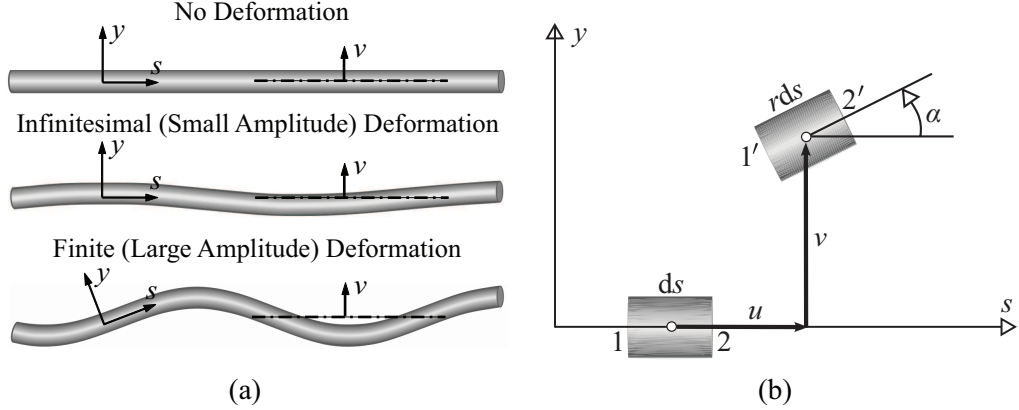}}
\caption{Kinematical representation of a planar flexural beam model: (a) infinitesimal versus finite deformation, (b) illustration of the variables $u$ and $v$ and the rotation angle $\alpha$, noting that $r=1+e$ where $e$ is the strain of the centerline.}
\label{BEM_deform}
\end{figure*}

The rod model admits longitudinal motion, tension or
compression, with $u$ denoting its axial displacements. In the
conventional beam model, the axial displacements are assumed to be
zero. Hence, the only variable is $v$, which denotes the lateral
displacements. The planar beam models admit both $u$ and $v$
displacements and provide a coupling relationship between these two
variables. Figure \ref{BEM_deform} provides graphical
descriptions of the kinematics of a planar beam showing the variables in the $s-y$ plane, where $s$ is the Lagrangian longitudinal coordinate. The third axis perpendicular to $s$ and $y$ is denoted by $z$. In a general planar beam model, the following kinematic relationship holds: $e=r-1$, where $e$ is the axial strain of the centerline and $r=\sqrt{(1+u')^2+v'^2}$. The inextensional planar beam model is a special case of the general planar beam model, in which an inextensionality constraint is applied such that the axial strain of the beam's centerline is assumed to be zero. This provides us with the following relationship:

\begin{equation}
u'=\sqrt{1-v'^2}-1.\label{Inex}
\end{equation}

\noindent Application of an inextensionality constraint to a beam with any type of boundary conditions is known to be generally adequate in the absence of large axial forces (Crespo da Silva and Glynn \cite{Silva}; Nayfeh and Pai \cite{Nayfeh_Pai}; Lacarbonara and Yabuno \cite{Lacar}). In all planar beam models, it is assumed that the cross-section remains plane and perpendicular to the centerline after elastic deformation. In addition, it is assumed that in-plane and out-of-plane warpings of the cross-section do not
occur. These assumptions are justified, even for large finite
deformation, in light of our focus on beams with aspect ratios high
enough for the Euler-Bernoulli kinematic hypthesis to be valid. In the spatial beam models, the beam admits four degrees of freedom, $u$, $v$, $w$, and $\gamma$, where $w$ denotes the lateral displacements in a direction perpendicular to
$u$ and $v$, and $\gamma$ is the torsion angle of the cross-section.
Here the displacement field is written in terms of the matrix
$R_{BS}$, which is defined as
\begin{equation*}
R_{BS}=\left[\begin{array}{ccc}\cos{\alpha}&-\sin{\alpha}&0\\
\sin{\alpha}&\cos{\alpha}&0\\
0&0&1\end{array}\right]\left[\begin{array}{ccc}\cos{\beta}&0&\sin{\beta}\\
0&1&0\\
-\sin{\beta}&0&\cos{\beta}
\end{array}\right]\left[\begin{array}{ccc}1&0&0\\
0&\cos{\gamma}&-\sin{\gamma}\\
0&\sin{\gamma}&\cos{\gamma}\end{array}\right]
\end{equation*}
in which $\alpha=\arctan{v'/h}$ and $\beta=-\arctan{w'/r}$, where $h$ is an agent variable defined as

\begin{align}
&h=1+u',
 \label{REM_h1}
\end{align}

\noindent and $r=\sqrt{h^2+v'^2}$. We note that $\alpha$ and $\beta$
are dependent variables representing the in-plane and out-of-plane
deflection angles, where the torsion angle $\gamma$ is an
independent variable. In a similar manner to the planar models, an
inextensionality constraint can be applied to produce the
inextensional spatial beam model. The advantage of the
inextensionality constraint is that it reduces the number of DOF of
a system by one, as indicated in Table \ref{Delta}. The cases we
consider in this work (rod, conventional beam and inextensional
planar beam) are all $1$ DOF systems. Future work will address finite-strain wave dispersion behavior of higher DOF systems (see
\cite{Abedinnasab_IMECHE_2012} for a static treatment of higher DOF
sytems).

\begin{flushleft}
\begin{table}
\caption{Rod and beam kinematic models}
\begin{tabular}{p{2.3cm}| c c c c}
\hline\hline Description&Motion/DOF&Indep. Variable(s)&Displacement
field, $\Delta$&Constraint\\\hline Rod&Longitudinal/1&$u$&$u$&---\\
Conventional beam&Transverse/1&$v$&\small$\left[\begin{array}{c}-yv'/\sqrt{1+v'^2}\\v-y(1-1/\sqrt{1+v'^2})\end{array}\right]$\normalsize&---\\
Inextensional planar beam&Planar/1&$u$,
$v$&\small$\left[\begin{array}{c}u-yv'/r\\v-y(r-h)/r\end{array}\right]$\normalsize&$r-1=0$\\
Planar beam&Planar/2&$u$,
$v$&\small$\left[\begin{array}{c}u-yv'/r\\v-y(r-h)/r\end{array}\right]$\normalsize&---\\
Inextensional spatial beam&Spatial/3&$u$, $v$, $w$,
$\gamma$&\footnotesize$\left[\begin{array}{c}u\\v\\w\end{array}\right]+\left(R_{BS}-I\right)\left[\begin{array}{c}0\\y\\z\end{array}\right]$\normalsize&$\sqrt{r^2+w'^2}-1=0$\\
Spatial beam&Spatial/4&$u$, $v$, $w$,
$\gamma$&\footnotesize$\left[\begin{array}{c}u\\v\\w\end{array}\right]+\left(R_{BS}-I\right)\left[\begin{array}{c}0\\y\\z\end{array}\right]$\normalsize&---\\\hline\hline
\end{tabular}
\label{Delta}
\end{table}
\end{flushleft}

\subsection{Finite strain fields of single DOF models}

In Section \ref{NEM}, we derive the equation of motion for our three
models on the basis of the Green-Lagrange strain tensor. In the following subsections, the form of this tensor is given, for a rod and for a beam. Other measures of strain may be chosen. However, in this work, we only use the Green-Lagrange strain owing to its wide use in the literature.

\subsubsection{Rod: Finite strain field}
Introducing $\Delta$ to generally represent elastic displacement, the Green-Lagrange strain field in a rod is given by

\begin{equation}
\epsilon=\frac{\partial\Delta}{\partial{}s}+\frac{1}{2}\left(\frac{\partial\Delta}{\partial{}s}\right)^2,\label{REM_strn1}
\end{equation}

\noindent where the first and second terms on the RHS represent the
linear and nonlinear parts, respectively. As seen in Table
\ref{Delta}, the elastic displacement field, $\Delta$, is scalar for a rod and
is equal to the axial displacement, $u$.


\subsubsection{Beam: Finite strain field}
In analogy to \eqref{REM_strn1}, the Green-Lagrange
strain tensor for a beam is

\begin{equation}
\epsilon_{ij}=\frac{1}{2}\left(\frac{\partial\Delta_i}{\partial{}x_j}+\frac{\partial\Delta_j}{\partial{}x_i}+\frac{\partial\Delta_m}{\partial{}x_i}\frac{\partial\Delta_m}{\partial{}x_j}\right),\label{BEM_strn}
\end{equation}

\noindent where $i$ and $j$ take the value of $1$ or $2$, in which
$x_1=s$ and $x_2=y$ (see Fig. \ref{BEM_deform} and
Table \ref{Delta}). For conventional and inextensional beams, $\Delta$ represents the elastic displacement vector, which is
$\left [-yv'/\sqrt{1+v'^2}~~  v-y(1-1/\sqrt{1+v'^2})\right]^{\rm T}$
and
$\left[ u-yv'/r ~~v-y(r-h)/r\right]^{\rm T}$, respectively.


\section{Equations of motion based on finite deformation}
\label{NEM}

In this section we derive the equation of motion for a rod (Section
\ref{REM}), a conventional beam (Section \ref{BEM}) and an
inextensional planar beam (Section \ref{IEM}). In our derivations, we
consider non-dissipative, isotropic media and assume
constant material and geometric properties. Furthermore, we ignore the effect of lateral inertia on the longitudinal motion in all models. 

\subsection{Rod: Longitudinal motion}
\label{REM}

Using Hamilton's principle, we write the equation of motion for a
rod under uniaxial stress as

\begin{equation}
\int_0^t {\left( {\delta T - \delta {U^e}+\delta W^{nc}}
\right)\textit{\emph{d}}t}  = 0,\label{REM_ham1}
\end{equation}

\noindent where $T$, $U^e$, and $W^{nc}$ denote kinetic energy,
elastic potential energy and work done by external non-conservative forces and
moments, respectively, and $t$ denotes time. Using integration by parts, the variation of kinetic energy is

\begin{align}
\delta T =&-\rho A\int_0^l(\ddot u\delta u)\textit{\emph{d}}s
.\label{REM_varT1}
\end{align}

\noindent where $l$ denotes the length of an arbitrary portion of the rod, and $\rho$ and $A$ denote densoty and cross-sectional area, respectively. The variation of elastic potential energy is also obtained using integration by parts and is given as

\begin{equation}
\delta{}U^e=\int_0^l\int_A(\sigma\delta\epsilon)\textit{\emph{d}}A\textit{\emph{d}}s,\label{REM_ue1}
\end{equation}

\noindent where $\sigma$ and $\epsilon$ are the axial stress and
axial strain, respectively. In this work we base our analysis on the Cauchy stress and consider a linear stress-strain relationship as given by Hooke's law, that is, $\sigma=E\epsilon$, where $E$ is the Young's modulus. Using Eq. \eqref{REM_ue1}, and with the aid of integration by parts, we can now write the variation of elastic potential energy as

\begin{equation}
\delta {U^e} = \int_0^l {\left\{ {\frac{1}{2}E A h(h^2-1)\delta u' }
\right\}} \textit{\emph{d}}s,\label{REM_cc1}
\end{equation}

\noindent where $u'={{\textit{\emph{d}} u}}/{\textit{\emph{d}}
s}=u_{s}$ and $h$ is as defined in Eq. \eqref{REM_h1}. The variation of the work done by non-conservative forces and moments is
given in terms of the variation of axial deformation, $u$, and the
distributed external axial load, $q_u$,

\begin{equation}
\delta {W^{nc}} = \int_0^l {\left({{q_u}\delta u}
\right)\textit{\emph{d}}s}. \label{REM_wnc1}
\end{equation}

\noindent Substitution of Eqs. \eqref{REM_varT1}, \eqref{REM_cc1} and \eqref{REM_wnc1} into Eq. \eqref{REM_ham1} produces the equation of
motion \eqref{REM_motion1} and the companion boundary conditions
\eqref{REM_boundary1}:

\begin{align}
&\int_0^t \left\{ {\int_0^l {\left( {{A_1}\delta u }
\right)\textit{\emph{d}}s+\left( {{B_1}\delta u
+{\overline{B_1}}\delta {u^\prime }} \right)\left|
\begin{array}{l} s = l \\s = 0 \\ \end{array}\right.}
}\right\}\textit{\emph{d}}t=0,\label{REM_motion1}
\end{align}

\begin{equation}
\left( {{B_1} = 0\quad \rm or\quad u = 0} \right)\quad \rm and\quad
\left( {{ \overline{B_1}} = 0\quad \rm or\quad {u^\prime } = 0}
\right). \label{REM_boundary1}
\end{equation}

\noindent We can now write an exact nonlinear equation of motion of
a 1D rod under finite deformation as

\begin{align}
A_1=0:\qquad\rho{A}\ddot{u}=\frac{1}{2}E
A(3h^2-1)u''+q_u.\label{REM_AEq1}
\end{align}

\noindent The section loads are

\begin{align}
&B_1=\frac{1}{2}E A h(h^2-1)\quad \rm and\quad\overline{B_1}=0,
\end{align}

\noindent where $B_1$ is the axial force. If the axial deformation is infinitesimal, then $u'$ is
small and from Eq. \eqref{REM_h1}, $h\approx1$. Substitution of
$h=1$ into Eq. \eqref{REM_AEq1} leads to

\begin{align}
A_1=0:\qquad\rho{A}\ddot{u}=E Au''+q_u.\label{REM_ALEq1}
\end{align}

\noindent which is the equation of motion describing infinitesimal
axial deformation.

\subsection{Conventional beam: Transverse motion}
\label{BEM}

Using Hamilton's principle, we state the equation of motion as follows:

\begin{equation}
\int_0^t {\left( {\delta T - \delta {U^g} - \delta {U^e}+\delta
W_{nc}} \right)\textit{\emph{d}}t}  = 0.\label{BEM_ham}
\end{equation}

\noindent We express the elastic deformation using the flexural
displacement $v$ as well as the angular displacement $\alpha$. These
two elastic coordinates are related by the following holonomic
constraint: $\alpha  = {{{\tan }^{ - 1}}({{\textit{\emph{d}}
v}}/{\textit{\emph{d}} s}}) = {\tan ^{ - 1}}{{v^\prime
}}$. The variation of kinetic energy, with the aid of
integration by parts, is

\begin{align}
\delta
T=&-\rho\int_0^l\{\{A\ddot{v}-\left(\frac{1}{r^2}J\ddot{\alpha}\right)'\}\delta v\}\textit{\emph{d}}s\nonumber\\
&-\rho\left\{\left(\frac{1}{r^2}J\ddot{\alpha}\right)\delta{}v\right\}\left|
\begin{array}{l} s = l \\s = 0 \\ \end{array}\right.,\label{BEM_varT}
\end{align}

\noindent where $J$ is the second moment of cross-sectional area,
defined as

\begin{align}
&J=\int_Ay^2\textit{\emph{d}}A=\int_Az^2\textit{\emph{d}}A=\frac{\pi}{4}a^4.\label{BEM_J}
\end{align}

\noindent In Eq. \eqref{BEM_varT}, $r$ is defined as
$r = \sqrt {1+{v^\prime}^2}$. The variation of gravitational potential energy is

\begin{equation}
\delta {U^g} = g\rho A\int_0^l {\left( {\delta v}
\right)\textit{\emph{d}}s}.\label{BEM_ug}
\end{equation}

\noindent where $g$ denotes acceleration due to gravity. As in the rod case, we use the Cauchy stress and the standard linear Hooke's law in our formulation. Since we are not permitting in-plane and out-of-plane warpings of the cross-section, we set the Poisson's ratio, $\nu$, to zero in the constitutive relationship. The variation of elastic potential energy can now be written as

\begin{equation}
\delta{}U^e=\int_0^l\int_A(\sigma_{ss}\delta\epsilon_{ss}+\sigma_{yy}\delta\epsilon_{yy}+2\sigma_{sy}\delta\epsilon_{sy})\textit{\emph{d}}A\textit{\emph{d}}s.\label{BEM_ue}
\end{equation}




\noindent After substituting Eq. \eqref{BEM_strn} and the constitutive relationship into \eqref{BEM_ue}, we obtain

\begin{equation}
\delta {U^e} = \int_0^l {\left\{ { C_{vp}\delta v' + C_{vpp}\delta
v''} \right\}} \textit{\emph{d}}s.\label{BEM_cc}
\end{equation}

\noindent where the coefficient of $\delta v'$ is

\begin{align}
C_{vp}=&\frac{1}{2}EA(v')^3+\frac{EJ(2-3r^2)v'(v'')^2}{2r^6}-\frac{EJ_fv'(v'')^4}{r^{10}},\label{BEM_Cvp1}
\end{align}

\noindent and the coefficient of $\delta v''$ is

\begin{equation}
C_{vpp}=\frac{EJ_f(v'')^3}{2r^8}-\frac{EJ(1-3r^2)v''}{2r^4}.\label{BEM_Cvpp1}
\end{equation}

\noindent In Eqs. \eqref{BEM_Cvp1} and \eqref{BEM_Cvpp1}, $J_f$ is
the $4^{th}$ moment of cross-sectional area, defined as

\begin{equation}
J_f=\int_Ay^4\textit{\emph{d}}A=\int_Az^4\textit{\emph{d}}A=\frac{\pi}{8}a^6.\label{BEM_Jf}
\end{equation}

\noindent It is also useful to define $J_c$, the higher-order
product of cross-sectional area,

\begin{equation}
J_c=\int_Ay^2z^2\textit{\emph{d}}A=\frac{\pi}{24}a^6.\label{BEM_Jc}
\end{equation}

\noindent Both $J_f$ and $J_c$ have been calculated for a circular
cross-section with radius $a$. It can be seen from Eqs.
\eqref{BEM_Jf} and \eqref{BEM_Jc} that for a circular cross-section
beam, $J_f=3J_c$. The variation
of the work done by non-conservative forces and moments is given in terms of the
variation of the flexural displacement and the distributed external
load,

\begin{equation}
\delta {W_{nc}} = \int_0^l {\left( {{q_v}\delta v}
\right)\textit{\emph{d}}s}.\label{BEM_wnc}
\end{equation}

\noindent Substitution of Eqs. \eqref{BEM_varT}, \eqref{BEM_ug}, \eqref{BEM_cc} and \eqref{BEM_wnc} into Eq. \eqref{BEM_ham} generates the equation of motion \eqref{NBEM_motion2} and the companion boundary conditions
\eqref{NBEM_boundary2},

\begin{align}
&\int\limits_0^t \left\{ {\int\limits_0^L {\left( {{A_2}\delta v }
\right)\textit{\emph{d}}s+\left( {{B_2}\delta v
+{\overline{B_2}}\delta {v^\prime }} \right)\left|
\begin{array}{l} s = L \\s = 0 \\ \end{array}\right.}
}\right\}\textit{\emph{d}}t=0,\label{NBEM_motion2}
\end{align}

\begin{equation}
\left( {{B_2} = 0\quad or\quad v = 0} \right)\quad \rm and\quad \left(
{{ \overline{B_2}} = 0\quad or\quad {v^\prime } = 0} \right).
\label{NBEM_boundary2}
\end{equation}

\noindent Consequently, an exact nonlinear equation of motion for a conventional
Euler-Bernoulli beam based on finite deformation is

\begin{align}
A_2=0:\qquad-\rho{A}g-\rho{A}\ddot{v}+\rho{}J\left(\frac{\ddot{\alpha}}{r^2}\right)'+q_v+(C_{vp})'-(C_{vpp})''=0.\label{BEM_Eq1}
\end{align}

\noindent After substituting the $C_{vp}$ and $C_{vpp}$ coefficients
form Eqs. \eqref{BEM_Cvp1} and \eqref{BEM_Cvpp1} and simplifying, we
obtain the closed-form equation of motion for a conventional beam,

\begin{align}
A_2=0:&-\rho{A}g-\rho{A}\ddot{v}+\rho{}J\left(\ddot{\alpha}\right)'+q_3+\frac{3EA(v')^2v''}{2}+\frac{2EJv'(1+3(v')^2)v''v^{(3)}}{\left(1+(v')^2\right)^3}\nonumber\\
&-\frac{EJ(2+3(v')^2)v^{(4)}}{2(1+(v')^2)^2}+\frac{EJ(1+4(v')^2-9(v')^4)(v'')^3}{2\left(1+(v')^2\right)^4}-\frac{3EJ_f(v'')^2v^{(4)}}{2\left(1+(v')^2\right)^4}\nonumber\\
&+\frac{24EJ_fv'(v'')^3v^{(3)}}{\left(1+(v')^2\right)^5}+\frac{3EJ_f(1-9(v')^2)(v'')^5}{\left(1+(v')^2\right)^6}-\frac{3EJ_fv''(v^{(3)})^2}{\left(1+(v')^2\right)^4}=0.\label{BEM_EqConv}
\end{align}

\noindent The section loads, namely the transverse shear force and
the bending moment, are respectively

\begin{align}
&B_2=\rho{}J\left(\frac{\ddot{\alpha}}{r^2}\right)+C_{vp}-(C_{vpp})'\quad \rm and\quad\overline{B_2}=C_{vpp}.\label{BEM_BCb}
\end{align}


\subsection{Inextensional planar beam: Longitudinal-transverse motion}
\label{IEM}

Following  Hamilton's principle, the inextensionality constraint as
stated in Eq. \eqref{Inex}, and its corresponding Lagrange
multiplier, $\vartheta$, we write

\begin{equation}
\int_0^t \left( \delta T - \delta {U^g} - \delta {U^e}+\delta
W_{nc}+\delta \left(\vartheta{}.e\right) \right)dt=0.\label{ham2}
\end{equation}

\noindent Accounting for rotary inertia, we write the total kinetic
energy of the beam as

\begin{equation}
T=1/2\rho{}\int_0^L\{A(\dot{u}^2+\dot{v}^2)+J\dot{\alpha}^2\}ds.\label{T}
\end{equation}

\noindent The variation of kinetic energy is then obtained with the
aid of integration by parts,

\begin{align}
\delta T=&-\rho\int_0^L\{\{A\left[\begin{array}{cc}\ddot u&\ddot
v\end{array}\right]-\left[\begin{array}{cc}-\frac{v'}{r^2}J\ddot{\alpha}&\frac{h}{r^2}J\ddot{\alpha}\end{array}\right]'\}\left[\begin{array}{c}\delta
u\\\delta v\end{array}\right]\}ds\nonumber\\
&-\rho\left[\begin{array}{cc}-\frac{v'}{r^2}J\ddot{\alpha}&\frac{h}{r^2}J\ddot{\alpha}\end{array}\right]\left[\begin{array}{c}\delta
u\\\delta v\end{array}\right]\left|
\begin{array}{l} s = l \\s = 0 \\ \end{array}\right..\label{varT}
\end{align}

\noindent Substituting Eq. \eqref{BEM_strn} and the linear constitutive relationship into Eq. \eqref{BEM_ue} yields the unknown coefficients of the
variation of elastic potential energy in Eq. \eqref{BEM_cc}. The
coefficient of $\delta v'$ in Eq. \eqref{BEM_cc} is

\begin{align}
C_{vp}=&-\left(J+\frac{J_f(1+2v'^2)v''^2)}{2(1-v'^2)}\right)\frac{Ev'v''^2}{(1-v'^2)^2}-\left(J+\frac{3v''^2}{2(1-v'^2)}\right)\frac{Ev'^2v^{(3)}}{1-v'^2}.\label{BEM_Cvp}
\end{align}

\noindent Similarly, the coefficient of $\delta v''$ in Eq.
\eqref{BEM_cc} is

\begin{equation}
C_{vpp}=\left(J+\frac{J_fv''^2}{2(1-v'^2)}\right)Ev''.\label{BEM_Cvpp}
\end{equation}

\noindent Substitution of Eqs. \eqref{varT}, \eqref{BEM_ug}, \eqref{BEM_cc} and \eqref{BEM_wnc} into Eq. \eqref{ham2}
generates the equation of motion \eqref{NNBEM_motion2} and the
companion boundary conditions \eqref{NNBEM_boundary2},

\begin{align}
&\int\limits_0^t \left\{ {\int\limits_0^L {\left( {{A_2}\delta v }
\right)\textit{\emph{d}}s+\left( {{B_2}\delta v
+{\overline{B_2}}\delta {v^\prime }} \right)\left|
\begin{array}{l} s = L \\s = 0 \\ \end{array}\right.}
}\right\}\textit{\emph{d}}t=0,\label{NNBEM_motion2}
\end{align}

\begin{equation}
\left( {{B_2} = 0\quad or\quad v = 0} \right)\quad \rm and\quad \left(
{{ \overline{B_2}} = 0\quad or\quad {v^\prime } = 0} \right).
\label{NNBEM_boundary2}
\end{equation}

\noindent Consequently, an exact nonlinear equation of motion for an inxetensional
Euler-Bernoulli beam based on finite deformation is

\begin{align}
A_2=0:\qquad-\rho{A}g-\rho{A}\ddot{v}+\rho{}J\left(\frac{\ddot{\alpha}}{r^2}\right)'+q_v+(C_{vp})'-(C_{vpp})''=0.\label{BEM_EqInex}
\end{align}

\noindent After substituting the $C_{vp}$ and $C_{vpp}$ coefficients
form Eqs. \eqref{BEM_Cvp} and \eqref{BEM_Cvpp} and simplifying, we
obtain the closed-form equation of motion for an inextensional
planar beam,

\begin{align}
A_2=0:&-\rho{A}g-\rho{A}\ddot{v}+\rho{}J\left(\frac{\ddot{\alpha}}{h}\right)'-\left[\frac{v'}{\sqrt{1-v'^2}}\left(\int_x^l(\rho{}A\ddot{u}-q_2)ds\right)\right]'+q_3\nonumber\\
&-\frac{EJv^{(4)}}{1-(v')^2}-\frac{4EJv'v''v^{(3)}}{\left(1-(v')^2\right)^2}-\frac{EJ(1+3(v')^2)(v'')^3}{\left(1-(v')^2\right)^3}-\frac{3EJ_f(v'')^2v^{(4)}}{2\left(1-(v')^2\right)^2}\nonumber\\
&-\frac{12EJ_fv'(v'')^3v^{(3)}}{\left(1-(v')^2\right)^3}-\frac{3EJ_f(1+5(v')^2)(v'')^5}{2\left(1-(v')^2\right)^4}-\frac{3EJ_fv''(v^{(3)})^2}{\left(1-(v')^2\right)^2}=0.\label{InexEq}
\end{align}

\noindent The transverse shear force and  bending moment section loads are found from Eqs. \eqref{InexBCa} and \eqref{InexBC}, respectively.

\begin{subequations}
\begin{align}
&B_2=\rho{}J\left(\frac{\ddot{\alpha}}{h}\right)-\left[\frac{v'}{\sqrt{1-v'^2}}\left(\int_x^l(\rho{}A\ddot{u}-q_2)ds\right)\right]\nonumber\\
&-\frac{EJv'(v'')^2}{\left(1-(v')^2\right)^2}-\frac{EJv^{(3)}}{1-(v')^2}-\frac{3EJ_f(v'')^2v^{(3)}}{2\left(1-(v')^2\right)}-\frac{3EJ_fv'(v'')^4}{2\left(1-(v')^2\right)^3},\label{InexBCa}
\end{align}
\begin{align}
&\overline{B_2}=EJv''+\frac{EJ_f(v'')^3}{2(1-(v'))^2}.\label{InexBC}
\end{align}
\end{subequations}

It is noticeable from Eqs. \eqref{BEM_Jf} and \eqref{BEM_J}, that
$J_f$ is quantitatively small compared to $J$ when the beam is
slender (i.e., when $a$ is small). As a result, we can neglect the
terms containing $J_f$ in Eqs. \eqref{InexEq}, \eqref{InexBCa} and \eqref{InexBC} (further discussion on this omission is provided in Section
\ref{No_Jf}). If these terms are dropped from Eq. \eqref{InexEq}
only three elastic terms remain, namely
$-\frac{EJv^{(4)}}{1-(v')^2}$,
$-\frac{4EJv'v''v^{(3)}}{\left(1-(v')^2\right)^2}$ and
$-\frac{EJ(1+3(v')^2)(v'')^3}{\left(1-(v')^2\right)^3}$. The
resulting  equation of motion is relatively simple yet still
adequate for large elastic deformation.

\section{Static deflection under finite deformation}

\begin{figure*}[!ht]
\centerline{\includegraphics{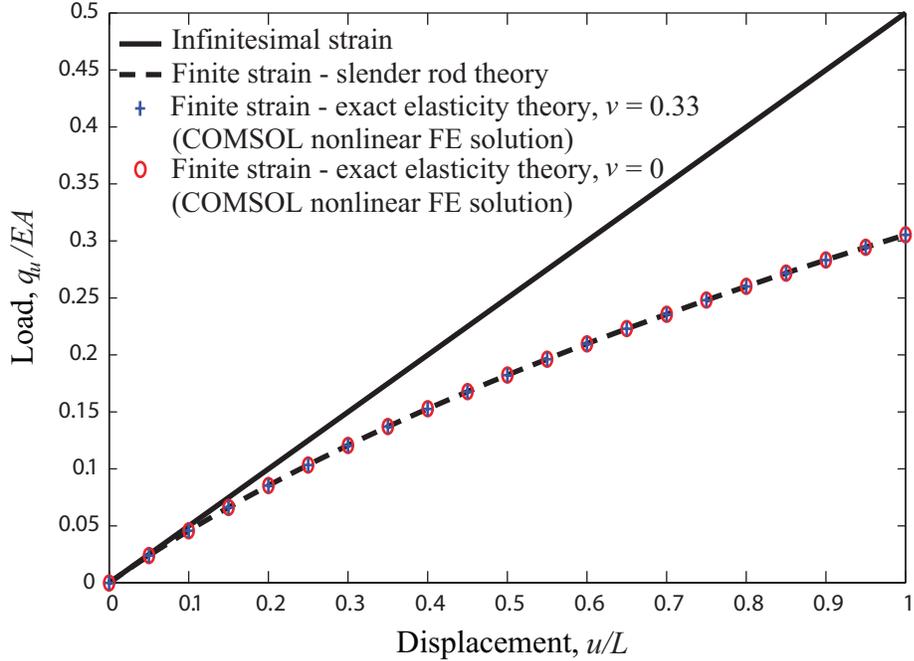}} \caption{Static
deflection of a rod.} \label{Rod_Comsol}
\end{figure*}

In this section we consider the static deflection problem and hence
omit all the inertial terms in the rod and beam equations of motion
(Eqs. \eqref{REM_AEq1}, \eqref{BEM_Eq1} and \eqref{InexEq}). Figure
\ref{Rod_Comsol} presents the static response of a rod due to the
application of a force load, $q_{u}$ (Section \ref{REM} ). We show
the response following our finite-deformation formulation and that
obtained using the COMSOL Multiphysics software package
\cite{Comsol} which is based on a three-dimensional (3D) nonlinear FE model of a rod following the theory of elasticity.\footnote[2]{The COMSOL model used is ``solid mechanics with geometric nonlinearity''. A fine mesh density (12,000 finite elements) was utilized and the calculations were stationary with a
maximum of 25 iterations in each step.} The results are in
excellent agreement and thus provide a validation to our rod
formulation for static response. Figure \ref{Rod_Comsol} also shows
that omission of the Poisson' ratio has a negligible effect on the
nonlinear static response. The deflection under the infinitesimal
strain assumption is included for comparison.

\begin{figure*}[!ht]
\centerline{\includegraphics{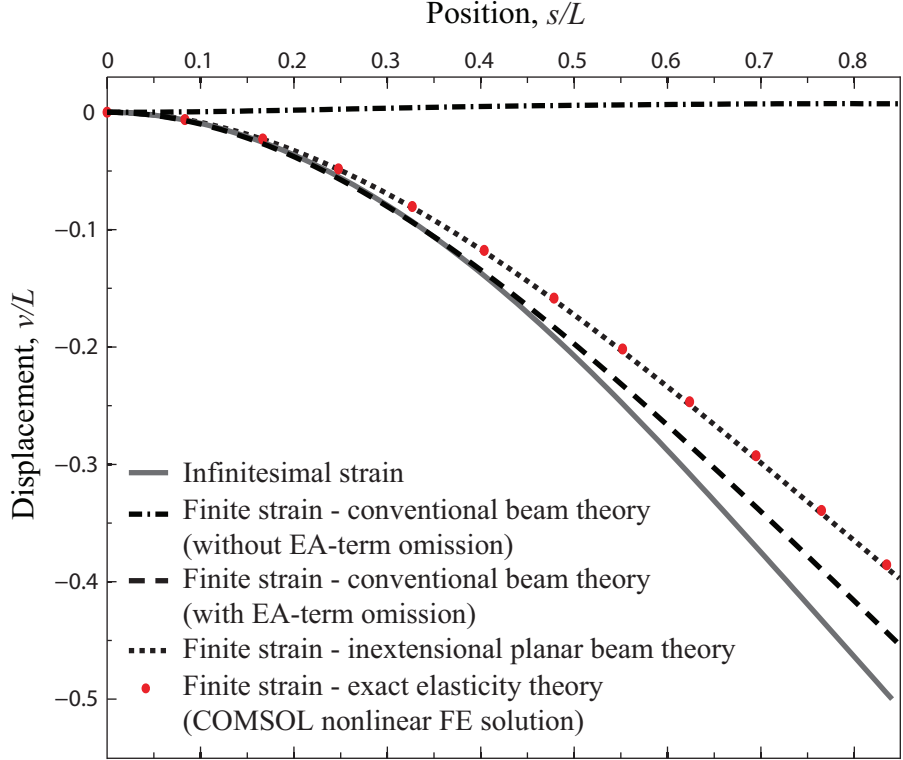}} \caption{Static
deflection of a clamped-free flexural beam
(${q_vL^3}/{8EJ}=-0.5$).} \label{Flex_EA}
\end{figure*}

In Fig. \ref{Flex_EA} we consider a clamped-free beam, with a length
of $L$, under a uniform external load, $q_v$. In the figure we show
the results obtained by our finite-deformation formulations
(following the conventional transverse motion model and the
inextensional planar motion model) and the result obtained by the
COMSOL Multiphysics software package (based on the same 3D nonlinear
FE model used for the rod). Once again, the deflection
under the infinitesimal strain assumption is included for
comparison.

It is clear that the inextensional planar beam theory model agrees
very well with the nonlinear FE solution based on the
theory of elasticity. This is a significant result considering that
the inextensional planar beam theory model is only a 1 DOF model.
The conventional beam theory model on the other hand perfoms poorly
and fails to even qualitatively capture the basic trend of the
response. Here we note that in the conventional beam theory model
there is a term that includes $''EA''$ (Eq. \eqref{BEM_Cvp1}). Yet,
in the inextensional planar beam theory model any term including
$''EA''$ drops out as a consequence of the derivations. Should we
forcefully drop the term that features $''EA''$ in Eq.
\eqref{BEM_Cvp1} we find that the response improves significantly
although still does not match the accuracy of the inextensional
planar motion model. For the wave propagation problem studied in
Sections \ref{NDR} and \ref{RD}, omission of the term including
``EA'' will also improve the performance of the conventional
transverse motion model although in that case the response will not
qualitatively follow the solution of the inextensional planar motion
model (and therefore this omission will not be illustrated in the
results we present in Section \ref{RD}).

\section{Wave dispersion under finite deformation}
\label{NDR}

In this section we derive exact amplitude-dependent dispersion
relations for harmonic wave propagation in a rod and in a flexural
beam under finite deformation. We also obtain the explicit frequency
versus wavenumber solutions - exactly for the rod problem and
approximately for the flexural beam problem.

\subsection{Rod: Exact analytical dispersion relation and solution}
\label{RBT}

Using Eq. \eqref{REM_h1}, we rewrite Eq. \eqref{REM_AEq1} as

\begin{align}
\ddot{u}-c_0^2u''=\frac{1}{2}\left[3c_0^2(u')^2+c_0^2(u')^3\right]',\label{RBT_eq}
\end{align}

\noindent where $c_0=\sqrt{{E}/{\rho}}$. Equation \eqref{RBT_eq}
is integrable despite it being nonlinear. Differentiating
\eqref{RBT_eq} once with respect to $s$ gives

\begin{align}
(\ddot{u})'-c_0^2u^{(3)}=\frac{1}{2}\left[3c_0^2(u')^2+c_0^2(u')^3\right]''.\label{RBT_eqp}
\end{align}

\noindent Defining $\bar{u}=u'$ and $\tau=\omega{}t$, where $\omega$
is the frequency of a harmonic wave, Eq. \eqref{RBT_eqp} becomes

\begin{align}
\omega{}^2\bar{u}_{\tau{}\tau}-c_0^2\bar{u}''=\frac{1}{2}\left[3c_0^2(\bar{u})^2+c_0^2(\bar{u})^3\right]''.\label{RBT_eqb}
\end{align}

\noindent Defining $z=\vert{}\kappa{}\vert{}s+\tau$, where $\kappa$
is the wavenumber of a harmonic wave, we rewrite Eq. \eqref{RBT_eqb}
as

\begin{align}
\omega{}^2\bar{u}_{zz}-c_0^2\kappa{}^2\bar{u}_{zz}=\frac{1}{2}\kappa{}^2\left[3c_0^2(\bar{u})^2+c_0^2(\bar{u})^3\right]_{zz},\label{RBT_eqz}
\end{align}

\noindent where now the explicit dependency on $s$ and $\tau$ has
been eliminated. Integrating Eq. \eqref{RBT_eqz} twice leads to

\begin{align}
\omega{}^2\bar{u}-c_0^2\kappa{}^2\bar{u}=\frac{1}{2}\kappa{}^2\left[3c_0^2(\bar{u})^2+c_0^2(\bar{u})^3\right],\label{RBT_eqii}
\end{align}

\noindent or,

\begin{align}
\left(\omega^2-c_0^2\kappa{}^2\right)\bar{u}-\frac{c_0^2\kappa{}^2}{2}\left(3\bar{u}^2+\bar{u}^3\right)=0.\label{RBT_feq}
\end{align}

\noindent We note that in our integartion of Eq. \eqref{RBT_eqz} we
get nonzero constants of integration in the form of polynomials in
$z$. Since these represent secular terms we have set them all equal
to zero in light of our interest in the dispersion relation.
Selecting the positive root of Eq. \eqref{RBT_feq} we get

\begin{align}
\bar{u}(z)=\frac{-3+\sqrt{1+8\omega{}^2/(c_0^2\kappa^2)}}{2}.\label{RBT_root}
\end{align}

\noindent Since $\bar{u}=u_s$, we recognize that
$\bar{u}=\vert{}\kappa{}\vert{}u_z$ and therefore Eq.
\eqref{RBT_root} represents a first order nonlinear ordinary
differential equation with $z$ and $u$ as the independent and
dependent variables, respectively.

Now we return to Eq. \eqref{RBT_eq} and consider for initial
conditions a sinusoidal displacement field, with amplitude $B$ and a
zero phase in time, and a zero velocity field. Following the change
of variables we have introduced, these initial conditions
essentially correspond to the following restrictions on the
$\bar{u}(z)$ function given in Eq. \eqref{RBT_root}:

\begin{equation}
\bar{u}(0)=\vert{}\kappa{}\vert{}B,\quad
\bar{u}_z(0)=0.\label{BBT_icu}
\end{equation}

\noindent We note that since $z$ describes a space-time wave phase,
the restrictions given in Eq. \eqref{BBT_icu} correspond to an
initial wave phase and hence may be viewed as ``initial conditions''
on the wave phase. The importance of these initial conditions is
that they incorporate the effect of the wave amplitude, $B$. Applying Eq.
\eqref{BBT_icu} to Eq. \eqref{RBT_root} allows us to use the latter
to solve for $\omega$ for a given value of $\kappa$ at $z=0$. Thus
we obtain an exact dispersion relation for wave propagation in a
rod under finite deformation, which is

\begin{align}
\omega(\kappa;B)=\sqrt{\frac{2+3B\vert{}\kappa{}\vert{}+(B\kappa{})^2}{2}}\omega_{\textit{\emph{inf}}},\label{RBT_fr}
\end{align}

\noindent where $\omega_{\textit{\emph{inf}}}$ is the frequency
based on infinitesimal deformation,

\begin{align}
\omega_{\textit{\emph{inf}}}(\kappa)=c_0\vert{}\kappa{}\vert{}.\label{RBT_fr_inf}
\end{align}

\noindent By taking the limit, $\mathop
{\textit{\emph{lim}}}\limits_{B \to 0}\omega(\kappa;B)$, in Eq.
\eqref{RBT_fr} we recover Eq. \eqref{RBT_fr_inf} which is the
standard linear dispersion relation for a rod.

\subsection{Conventional beam: Exact analytical dispersion relation}
\label{ConBeam}
Differentiating Eq. \eqref{BEM_Eq1} once with respect to $s$ and omitting the
external loads and gravity yields

\begin{align}
-\rho{A}(\ddot{v})'+\rho{}J\left(\frac{\ddot{\alpha}}{r^2}\right)''+(C_{vp})''-(C_{vpp})'''=0.\label{BBT_EqD}
\end{align}

\noindent Defining $\bar{v}=v'$, $\tau=\omega{}t$ and
$z=\kappa{}s+\tau$,\footnote[3]{From this point onwards, we replace $\vert{}\kappa{}\vert{}$ with $\kappa$ since only even-ordered functions of the wavenumber appear in the derived equations of motion.} we re-write Eq. \eqref{BBT_EqD} as

\begin{align}
-\rho{A}\omega^2(\bar{v}_{zz})+\rho{}J\kappa{}^2\omega^2\left(\frac{\alpha{}_{zz}}{r^2}\right)_{zz}+\kappa{}^2(C_{vp})_{zz}-\kappa{}^3(C_{vpp})_{zzz}=0.\label{BBT_EqD2}
\end{align}

\noindent Integrating Eq, \eqref{BBT_EqD2} twice with respect to $z$
gives us

\begin{align}
-\rho{A}\omega^2\bar{v}+\rho{}J\kappa{}^2\omega^2\frac{\alpha{}_{zz}}{r^2}+\kappa{}^2C_{vp}-\kappa{}^3(C_{vpp})_{z}=0,\label{BBT_EqI}
\end{align}

\noindent where $\alpha=\textit{\emph{tan}}^{-1}{\bar{v}}$ and

\begin{subequations}
\begin{align}
&C_{vp}=1/2EA(\bar{v})^3+EJ\kappa{}^2\frac{(2-3r^2)\bar{v}(\bar{v}_z)^2}{2r^6}-EJ_f\kappa{}^4\frac{\bar{v}(\bar{v}_z)^4}{r^{10}},\label{BBT_Cvpa}
\end{align}
\begin{align}
&C_{vpp}=EJ_f\kappa{}^3\frac{(\bar{v}_z)^3}{2r^8}-EJ\kappa{}\frac{(1-3r^2)\bar{v}_z}{2r^4},\label{BBT_Cvpb}
\end{align}
\begin{align}
&r=\sqrt{1+(\bar{v})^2}.\label{BBT_Cvpc}
\end{align}
\end{subequations}



\noindent In deriving Eq. \eqref{BBT_EqI} we set the constants of integration to zero and thus drop out the secular terms as we did for the rod model.
Furthermore, we recognize that $\bar{v}=v_s$ and therefore
$\bar{v}=\vert{}\kappa{}\vert{}v_z$.  In an analogous manner to the
rod problem (see Sec. \ref{RBT}), we write the following initial
conditions for the wave phase:

\begin{equation}
\bar{v}(0)=B\kappa{},\quad
\bar{v}_z(0)=0.\label{BBT_ic}
\end{equation}

\noindent Eq. \eqref{BBT_EqI}, which is a function of $B$, can be
solved numerically using a standard root finding technique to obtain
the explicit dispersion relationship between frequency and
wavenumber for a conventional beam under finite deformation.

\subsection{Inextensional planar beam: Exact analytical dispersion relation}

Differentiating Eq. \eqref{BEM_EqInex} once with respect to $s$ and
omitting the external loads and gravity yields

\begin{align}
-\rho{A}(\ddot{v})'+\rho{}J\left(\frac{\ddot{\alpha}}{r^2}\right)''+(C_{vp})''-(C_{vpp})'''=0.\label{BBT_EqDInex}
\end{align}

\noindent Defining $\bar{v}=v'$, $\tau=\omega{}t$ and
$z=\kappa{}s+\tau$, Eq. \eqref{BBT_EqDInex} is
re-written as

\begin{align}
-\rho{A}\omega^2(\bar{v}_{zz})+\rho{}J\kappa{}^2\omega^2\left(\frac{\alpha{}_{zz}}{r^2}\right)_{zz}+\kappa{}^2(C_{vp})_{zz}-\kappa{}^3(C_{vpp})_{zzz}=0.\label{BBT_EqD2Inex}
\end{align}

\noindent Integrating Eq. \eqref{BBT_EqD2Inex} twice leads to

\begin{align}
-\rho{A}\omega^2\bar{v}+\rho{}J\kappa{}^2\omega^2\frac{\alpha{}_{zz}}{r^2}+\kappa{}^2C_{vp}-\kappa{}^3(C_{vpp})_{z}=0,\label{BBT_EqIInex}
\end{align}

\noindent where $\alpha=\textit{\emph{tan}}^{-1}{\bar{v}}$ and

\begin{subequations}
\begin{align}
C_{vp}=&-\left(J+\frac{\kappa^2J_f(1+2\bar{v}^2)\bar{v}_z^2)}{2(1-\bar{v}^2)}\right)\frac{E\kappa^2\bar{v}\bar{v}_z^2}{(1-\bar{v}^2)^2}-\left(J+\frac{3\bar{v}_z^2}{2(1-\bar{v}^2)}\right)\frac{E\kappa^2\bar{v}^2\bar{v}_{zz}}{1-\bar{v}^2},\label{BEM_CvpInex}
\end{align}
\begin{align}
C_{vpp}=&\left(J+\frac{J_f\kappa^2\bar{v}_z^2}{2(1-\bar{v}^2)}\right)E\kappa{}\bar{v}_z.\label{BEM_CvppInex}
\end{align}
\end{subequations}

\noindent In deriving Eq. \eqref{BBT_EqIInex} we set the
constants of integration to zero in order to drop out the secular
terms as we did for the rod and conventional beam models.
Furthermore, and as we stated for the conventional beam model,
$\bar{v}=v_s$ and therefore $\bar{v}=\vert{}\kappa{}\vert{}v_z$.
Following the same approach as in Sections \ref{RBT} and
\ref{ConBeam}, we use the initial conditions stated in Eq. \eqref{BBT_ic} for the wave phase. Subsequently, Eq. \eqref{BBT_EqIInex}, which is a function of $B$, can be solved numerically using a standard root finding technique to
obtain the explicit dispersion relationship between frequency and
wavenumber for an inextensional planar beam under finite deformation.


\section{Results and discussion}
\label{RD}

\subsection{Finite-deformation dispersion curves}

\begin{figure*}[!ht]
\centerline{\includegraphics{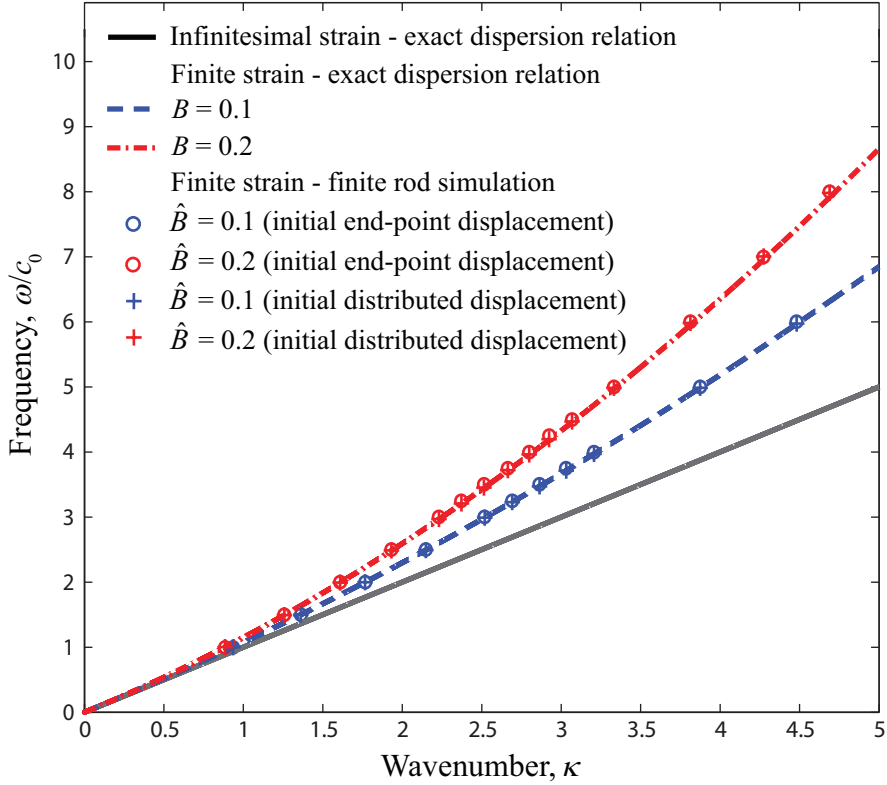}}
\caption{Frequency dispersion curves for a rod. The finite-strain
dispersion relation is based on Eq. \eqref{RBT_fr}; the
infinitesimal strain dispersion relation is based on Eq.
\eqref{RBT_fr_inf}.} \label{Axial_Dispersion_Curve}
\end{figure*}

Two amplitude-dependent finite-deformation dispersion curves for an
infinite rod based on Eq. \eqref{RBT_fr} are plotted in Fig.
\ref{Axial_Dispersion_Curve}. These dispersion curves provide an
exact fundamental description of how an elastic harmonic wave
locally, and instantaneously, disperses in an infinite rod under the
dynamic condition of amplitude-dependent finite deformation. Superimposed in
the same figure is the dispersion curve based on infinitesimal
deformation, i.e., Eq. \eqref{RBT_fr_inf}. We observe that the
deviation between a finite-deformation curve and the infinitesimal-deformation curve increases gradually with wavenumer, and the effect of the
wave amplitude on this deviation appears to be relatively steady - in the
wavenumber range considered - as $B$ is increased. More thorough
inspection of this deviation is provided in Section
\ref{CompInfFin}.

In addition, the results from standard FE simulations of
a finite version of the rod with length $L$ are presented. A
prescribed sinusoidal displacement with frequency $\hat{\omega}$ and
amplitude $\hat{B}$, i.e. $u(L,t)=\hat{B}\sin{\hat{\omega}t}$, was
applied to the tip of the rod with free-free boundary conditions.
The finite-deformation FE model consisted of 60
piecewise linear elements with equal lengths, and each node
consisted of two degrees of freedom, $u$ and $u'$. Equal time steps
of $10^{-4}$ [s] were considered in the numerical integration which
was implemented using MATLAB's $ode113$ solver \citep{Matlab}. The
wavenumber has been recorded by observing the wavelength after one
period of temporal oscillation of the tip (i.e., excitation point) of
the rod, and plotted as a function of frequency $\hat{\omega}$ for
two given amplitudes. This recording was possible because at the
vicinity of the excitation point the wave's harmonic form was
effectively still maintained during the first oscillation cycle. The
data points from this simulation (of a finite rod) match very well
with the analytically derived exact dispersion curve (corresponding
to an infinite rod). While the wave considered at the tip of the
excited rod will evolve, under finite strain, into a complex form as
it propagates into the rod, this correlation provides a validation
that a given harmonic wave will locally and instantaneously disperse in a manner exactly as described by Eq. \eqref{RBT_fr}. A second set of independent simulation data is also shown in the figure. This set corresponds to the response due to initial sinusoidal displacements at a prescribed wavenumber (applied at a state when the rod is at rest). Here we measured the frequency of  oscillation as the wave propagates considering the first temporal cycle. The simulation parameters are the same to those used in the runs where an initial end-point displacement is prescribed. As illustrated in the figure, this simulation provides yet another validation of the analytical dispersion relation given by Eq. \eqref{RBT_fr}.

\begin{figure*}[!ht]
\centerline{\includegraphics{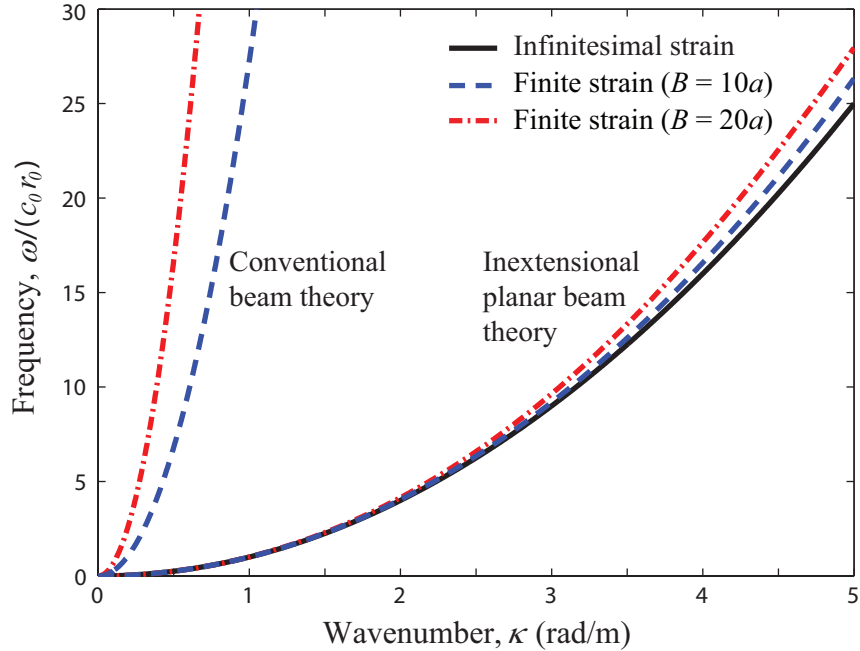}}
\caption{Frequency dispersion curves for a conventional beam and an
inextensional planar beam. The finite-strain curves represent exact
solutions of Eqs. \eqref{BBT_EqI} and \eqref{BBT_EqIInex},
respectively, evaluated by numerical root-finding. The infinitesimal-strain curve (based on Eq. \eqref{BBT_lfr}) is also
shown.}\label{Flex_Dispersion_Curve}
\end{figure*}

In Fig. \ref{Flex_Dispersion_Curve}, we present dispersion curves
for the conventional and inextensional planar beam models. For each
of the models, we show results for two different amplitudes, $B$. In
the interest of generality, the amplitude is considered in relation
to the beam's cross-sectional radius, $a$. The explicit relationship
between frequency and wavenumber is obtained by numerically solving
for the roots of Eq. \eqref{BBT_EqI} for the conventional beam model
and the roots of Eq. \eqref{BBT_EqIInex} for the inextensional
planar beam model.  Also shown in Fig. \ref{Flex_Dispersion_Curve}
is the dispersion curve based on infinitesimal deformation, which is

\begin{align}
\omega_{\textit{\emph{inf}}}(\kappa)=\frac{c_0r_0\kappa{}^2}{\sqrt{1+r_0^2\kappa{}^2}}.\label{BBT_lfr}
\end{align}

\noindent The finite-strain curve of the conventional beam model
shows a rapid change in slope at low frequencies compared to the
infinitismal-strain curve, which suggests poor peformance as was
seen in the static case. The inextentional beam response on the
other hand shows asymptotic convergence with the linear
dispersion as $\omega \rightarrow 0$ and the deviation grows
slowly with $\kappa$. We can also observe the effect of the wave
amplitude on the disperison: as $B$ increases the dispersion curve
rises at an increasing rate. Further comparison between the finite-deformation and infinitesimal-deformation responses is provided in
Section \ref{CompInfFin}.

\begin{figure*}[!ht]
\centerline{\includegraphics{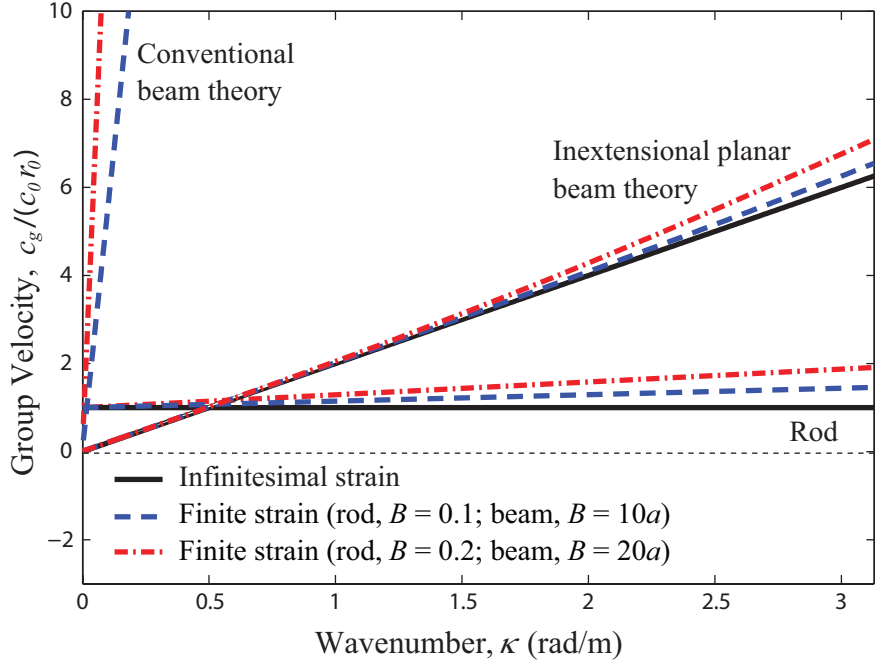}}
\caption{Group velocity dispersion curves for a rod, a conventional
beam and an inextensional planar beam under finite strain.}\label{Flex_Axial_Group_Velocity}
\end{figure*}

Figure \ref{Flex_Axial_Group_Velocity} shows the group velocity versus
wavenumber curves for the same cases considered in Figs.
\ref{Axial_Dispersion_Curve} and \ref{Flex_Dispersion_Curve}. The
group velocity is defined as
$c_{g}={\partial\omega}/{\partial\kappa}$. The figure puts on
view the significant qualitative transformation that emerges when
finite deformation is incorporated in wave propagation analysis. For
axial waves the group velocity, which is otherwise constant, is now
varying with $\kappa$. And for flexural waves the group velocity
follows a nonlinear function with $\kappa$ in contrast to the linear
trend for the infinitesimal-deformation model. Needless to say, the
intensity of these finite-deformation effects on the group velocity
increases with $B$ as mentioned in our discussion of the frequency
dispersion curves. A phase velocity dispersion diagram may also be generated, and from Figs. \ref{Axial_Dispersion_Curve} and \ref{Flex_Dispersion_Curve} it is clear that the trends will be similar to Fig. \ref{Flex_Axial_Group_Velocity}.

The dispersion characteristics presented provide an elucidation of
the nonlinear, finite-deformation dynamics of rods and beams
particularly for long structures or when the focus is on high
frequency/short wavelength behavior where a standard analysis based
on superposition of standing waves is generally not effective
\citep{BehbahaniPerkins_1996}. From a conceptual point of view, an increase in group velocities as seen in Fig. \ref{Flex_Axial_Group_Velocity} is particularly
significant in the study of waves where the rate and intensity of
energy propagation is consequential, such as in seismic waves
\cite{Sen_1991,Ostrovsky_2001}. It is also relevant to the study of
the speed of propagation of dislocations and cracks considering that large amplitudes have been observed near slip planes and crack tips
\cite{Gumbsch_1999,RosakisAJ_1999} (recalling that the phase and group velocities change with deformation amplitude). Future work will introduce appropriate extensions in the implementation of our methodology to address the specific geometric and material characteristics of these applications.

From an engineering design perspective, the amplitude-dependent
dispersion behavior we observe in Figs.
\ref{Axial_Dispersion_Curve}, \ref{Flex_Dispersion_Curve} and
\ref{Flex_Axial_Group_Velocity} potentially could be of benefit to
numerous applications; for example, the frequency shifts observed
suggest possible utilization of regular rods and beams as
amplitude-dependent wave propagation filters. Further utilization is
possible when these properties are considered in conjunction with
other avenues for design such as the introduction of periodicity
\cite{HusseinJSV06,HusseinJSV07,LiuL_JAM_2012}. Here we note that the derived finite-strain dispersion relations can be incorporated directly into transfer matrix models of 1D periodic media, which will be the focus of future research.


\subsection{Comparison with infinitesimal-deformation dispersion curves}
\label{CompInfFin}

\begin{figure*}[!ht]
\centerline{\includegraphics{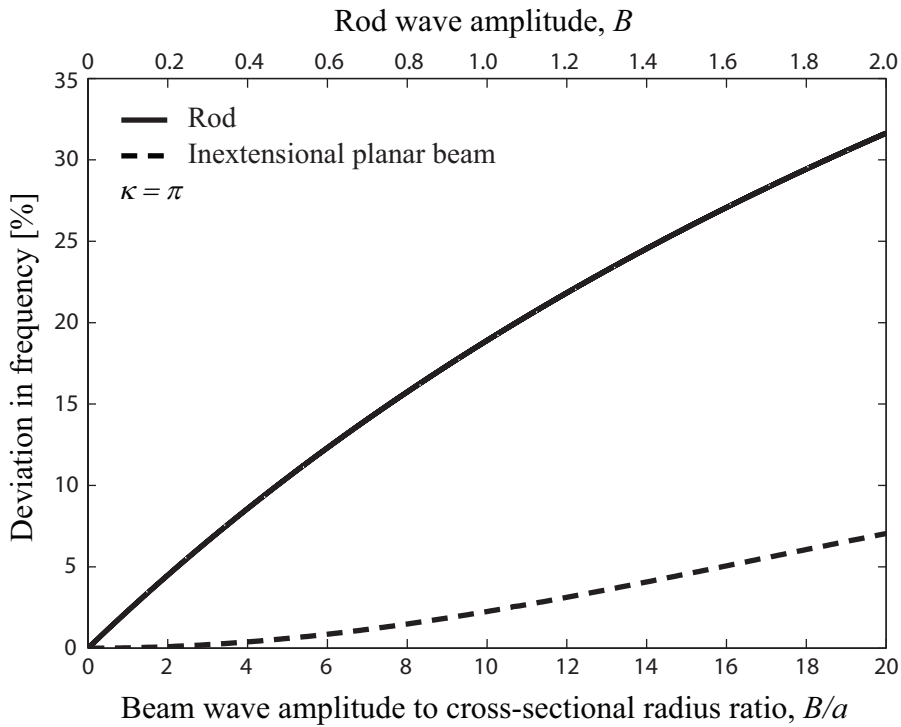}} \caption{Percentage
deviation in frequency when the dispersion curves are based on
infinitesimal strain versus finite strain (evaluated at
$\kappa=\pi$).} \label{Error}
\end{figure*}

Figure \ref{Error} presents the percentage deviation in the dispersion
curves when finite deformation is incorporated compared to when
infinitesimal deformation is assumed (evaluated at $\kappa=\pi$).
Clearly the higher the wave amplitude, the higher the deviation. For
a rod, the deviation follows a concave curve (increases at a
decreasing rate); whereas for an inextensional planar beam it
follows a convex curve (increases at an increasing rate). The level
of deviation for a given wave amplitude is considerably higher for
longitudinal waves in a rod compared to longitudinal-transverse waves in the
inextensional planar beam.

\subsection{Effect of the $4^{th}$ moment of area, $J_f$}
\label{No_Jf}

\begin{figure*}[!ht]
\centerline{\includegraphics{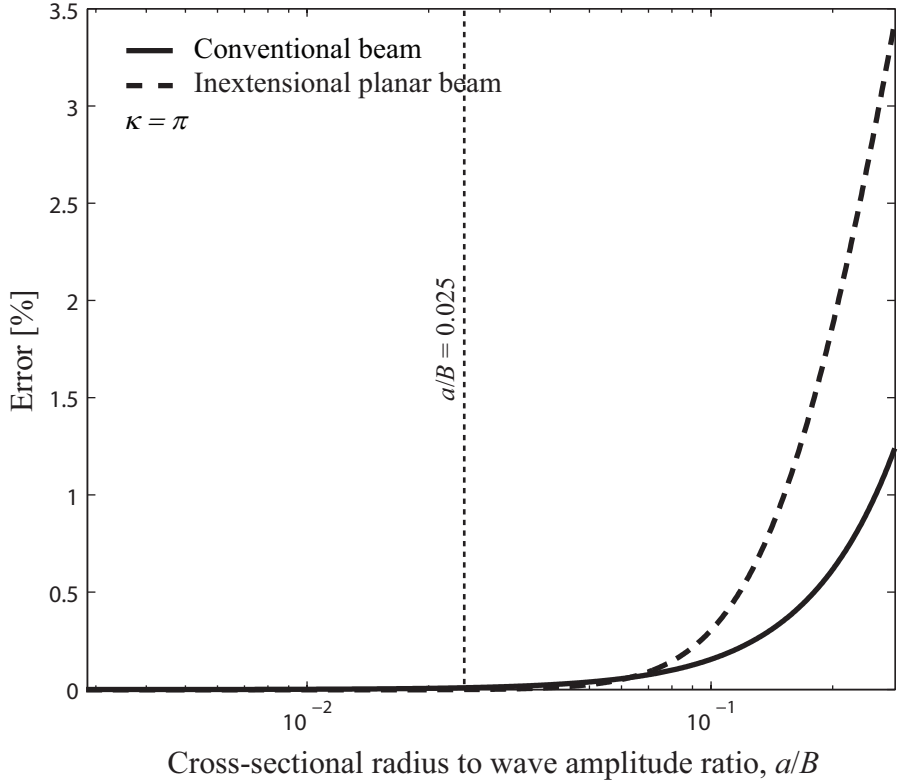}} \caption{Percentage
error in frequency due to omission of terms including $J_f$ in Eqs.
\eqref{BEM_Eq1} and \eqref{InexEq} for conventional and
inextensional planar beams (evaluated at $\kappa=\pi$).}
\label{Error_Jf}
\end{figure*}

In this section we examine the effect of the $4^{th}$ moment of
cross-sectional area, $J_f$, on the finite-deformation dispersion
relations for flexural beams. If, by way of example, we consider a
beam with a circular cross section of radius $a$, then the terms
$r_0$ and $r_f$ would each be a function of $a$, i.e., $r_0=\sqrt{{J}/{A}}={a}/{2}$ and $r_f=\sqrt{{J_f}/{A}}={a^2}/{2\sqrt{2}}$. Hence we get $r_f/r_0={a}/{\sqrt{2}}$ from which we can deduce that any term that contains $J_f$ diminishes when the cross-sectional area is small. Fig. \ref{Error_Jf} shows the effect of omission of all terms
that contain $J_f$ in both the conventional and inextensional planar
beam models. From the figure we note that for a ratio of
cross-sectional radius to wave amplitude, $a/B$, smaller than
approximately $0.025$, the effect of omitting the terms that contain
$J_f$ in Eqs. \eqref{BEM_EqConv} and \eqref{InexEq} is negligible.
However, for higher values of this ratio, the error introduced by
this omission increases dramatically.


\section{Conclusions}

We have derived and solved an exact dispersion relation for finite-strain wave propagation in a rod and an Euler-Bernoulli beam. This represents the first derivation of amplitude-dependent dispersion relations for elastic media under finite deformation, providing an analogy to the derivation of anharmonic (nonlinear) dispersion relations of crystals in condensed matter physics.

For axial motion (i.e., the rod problem), the closed-form exact frequency versus wavenumber solution was derived as a function of wave amplitude. The solution matches very well with data points obtained from high-resolution
FE simulations of a finite rod under finite deformation. For flexural motion (i.e., the conventional and inextensional planar beam problems), an approximate solution was obtained by numerically solving for the roots of the derived exact dispersion relation. A key aspect of the derivations that lead to the dispersion relation for each of these problems involves the introduction of a space-time wave phase, $z$, and the development of a strain relationship which at $z = 0$, with the appropriate initial conditions, gives the amplitude-dependent dispersion relation.

The results show that finite deformation causes the frequency (and therefore the phase and group velocity) dispersion curves to shift for both axial and
flexural waves in comparison to when infinitesimal deformation is assumed.  The level at which these nonlinear effects take place is dependent on the amplitude of the travelling wave, and the degree of shift in the disperison curves increases with wavenumber. Qualitatively, finite deformation is shown to introduce dispersion to an otherwise nondispersive medium. Both the rod and inextensional beam solutions display asymptotic convergence to the corresponding linear, infinitesimal-strain solution at low frequencies and at low wave amplitudes. For our choice of Cauchy stress and Green-Lagrange strain, all observed shifts are positive, i.e., the curves rise with finite deformation. Future work will examine the response based on other stress and strain measures, which in itself will serve as a rigorous characterization of the spectral properties of these measures. The results also show, for wave dispersion as well as static deflection, that a coupling between the longitudinal and transverse motions is necessary for an accurate description of the response.

Our analysis provide a means for elucidating the nonlinear dynamics
of rods and flexural beams particularly for long structures or when the focus
is on high frequency/short wavelength behavior. It also has implications on the study of wave phenomena in a broader range of problems where large elastic motion takes place. Examples where large deformations have been experimentally observed include waves produced by cracks and dislocations (problems of great importance in seismology, fracture mechanics and plasticity). In these problems, the standard dispersion relations based on infinitesimal motion cannot predict any alterations in the speed of sound due to the intensity of the motion, and hence the need for a finite-strain treatment as we observed in Fig. \ref{Flex_Axial_Group_Velocity} for the 1D models considered. While our present formulations consider linear constitutive relations (in order to isolate the effect of finite motion), the proposed methodology is applicable to problems that also exhibit material nonlinearity - which would be necesary for the applications mentioned above. From an engineering design perspective, the amplitude-dependent shifts in frequency, phase and group velocity in the dispersion spectrum provide promising avenues for the development of novel materials, structures and devices.

The presented analytical formalism for dispersion analysis under
finite deformation lays the foundation for systematic extensions to
more complex 1D models (e.g., rods/beams incorporating lateral inertia in
longitudinal motion, beams incorporating corotational motion, higher DOF beams) as well as multi-dimensional models (e.g., thin plates and shells, 2D plain stress/strain models, semi-infinite surfaces, 3D rods and plates, bulk media).


\bibliographystyle{model1-num-names}
\bibliography{rfs}

\end{document}